\documentclass[aps,prl,reprint,superscriptaddress]{revtex4-1}
\usepackage{amsmath,amssymb}
\usepackage[colorlinks=true,allcolors=blue]{hyperref}
\usepackage{dcolumn}
\usepackage{units}
\usepackage{graphicx}
\usepackage{float}
\usepackage{enumitem}
\usepackage{diagbox}
\usepackage{overpic}

\allowdisplaybreaks[1]

\renewcommand{\d}{\text{d}}
\newcommand{\e}{{\textbf{e}}}
\renewcommand{\r}{{\textbf{r}}}
\newcommand{\ee}{{\text{e}}}
\newcommand{\rr}{{\text{r}}}

\begin{document}

\title{Derivative of the light frequency shift as a measure of spacetime curvature \\ for gravitational wave detection}

\author{Giuseppe~Congedo}
\email{giuseppe.congedo@physics.ox.ac.uk}
\affiliation{Department of Physics, University of Oxford, Keble Road, Oxford OX1 3RH, United Kingdom}

\date{\today}


\begin{abstract}
The measurement of frequency shifts for light beams exchanged between two test masses nearly in free fall is at the heart of gravitational wave detection. It is envisaged that the derivative of the frequency shift is in fact limited by differential forces acting on those test masses.
We calculate the derivative of the frequency shift with a fully covariant, gauge-independent and coordinate-free method. This method is general and does not require a congruence of nearby beams' null geodesics as done in previous work.
We show that the derivative of the parallel transport is the only means by which gravitational effects shows up in the frequency shift. This contribution is given as an integral of the Riemann tensor --the only physical observable of curvature-- along the beam's geodesic. The remaining contributions are: the difference of velocities, the difference of non-gravitational forces, and finally fictitious forces, either locally at the test masses or non-locally integrated along the beam's geodesic. As an application relevant to gravitational wave detection, we work out the frequency shift in the local Lorentz frame of nearby geodesics.
\end{abstract}


\maketitle

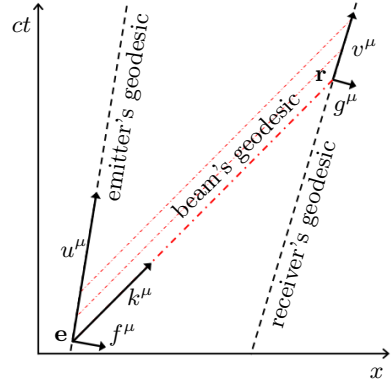
\begin{figure}[h!]
\centering
\begin{overpic}[width=0.45\textwidth]{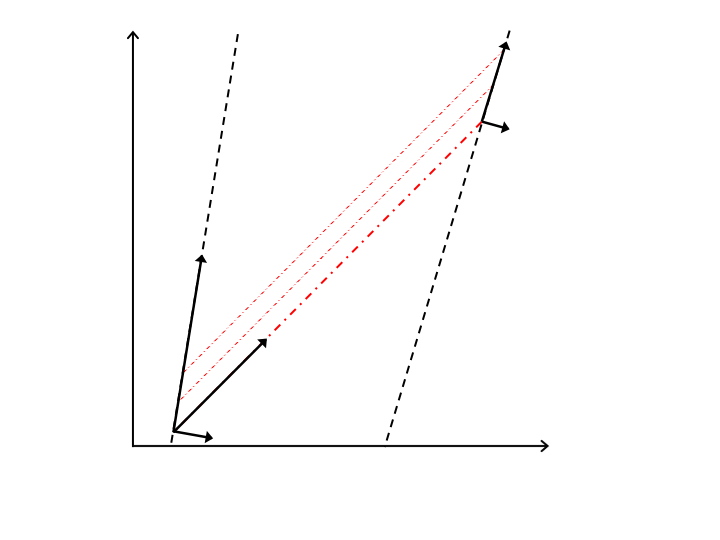} 
\put (14,66) {$ct$}
\put (73,9) {$x$}
\put (22,29) {$u^\mu$}
\put (70,62) {$v^\mu$}
\put (33,21) {$k^\mu$}
\put (21,15) {$\e$}
\put (64,58) {$\r$}
\put (67,53) {$g^\mu$}
\put (30,15) {$f^\mu$}
\put (29,38) {\rotatebox{81}{emitter's geodesic}}
\put (56,17) {\rotatebox{74}{receiver's geodesic}}
\put (40,35) {\rotatebox{45}{beam's geodesic}}
\end{overpic}
\vspace{-20pt}
\caption{\footnotesize{Instantaneous Minkowski diagram for the thought experiment of two test masses exchanging light beams and measuring the corresponding frequency shift. The 45$^\circ$ dot-dashed line is the null geodesic connecting the emission event $\e$ to the reception event $\r$. Null geodesics at later instants are shown as thinner lines. The other dashed lines (with slopes $>45^\circ$) are the two time-like emitter and receiver's geodesics. 4-velocities, 4-forces and the beam's 4-momentum are also displayed for clarity. The derivative of the fractional frequency shift depends on the difference between tensors, including the Riemann tensor and fictitious forces, at $\r$ and $\e$, the latter being delayed by the beam's light time along the null geodesic.}}
\label{fig:diagram}
\end{figure}

\textit{Introduction}-- The exchange of light beams between (almost) free falling test masses and the measurement of the corresponding frequency shifts (see Fig.\;\ref{fig:diagram}) is at the heart of any thought (real) experiment devised for measuring in principle (in practice) space-time curvature. It is in fact the key for the direct observation of gravitational waves (GW) by interferometer detectors on the ground \cite{harry2010,arcenese2015} and in space \cite{amaro2012}, and pulsar timing arrays \cite{hoobs2010}. 
Particularly at low frequency, GW detectors are limited by differential forces acting on the test masses \cite{congedo2015} and, as such, the derivative of the frequency shift may be a good observable of space-time curvature and, in general, a means to separate true gravitational forces from spurious effects. Recently, two different approaches \cite{koop2014,congedo2013} have put forward a formalism that allows the frequency shift to be computed in terms of an integrated measure of curvature. This is distinct from earlier attempts that often relied on simplifying assumptions, e.g.\ metric expansion, geodesic deviation, choice of a preferred coordinate system, or fixing an \textit{a priori} gauge (see the introduction of Ref.\ \cite{koop2014} and references therein).

The only physical covariant quantity that unambiguously describes the effect of curvature in vacuum is the Riemann tensor \footnote{One may argue that the Weyl tensor could be a more general curvature tensor, although it reduces to the Riemann tensor in vacuum.}. In effect, other general relativistic variables, such as the Ricci tensor and the Ricci scalar are identically zero in vacuum, even in a curved space-time. Additionally, the Christoffel symbols can be set to zero by a proper change of reference frame and the metric itself is in general gauge dependent \cite{misner}. Consequently, all those quantities are not good observables of the true gravitational effect and, as such, they might eventually lead to ambiguous results. Previous work has already pointed out, although in different formulations, that the frequency shift is sensitive to an integrated measure of the Riemann tensor over the space-time between the two test masses. Those formulations worked out this contribution by defining either a null congruence of present and past null geodesics \cite{koop2014}, or a non-standard time-like congruence of the emitter's velocity and the receiver's velocity \cite{congedo2013}. Although the relation between those two approaches is not fully clear at the moment, the results are similar.

In this work we give a covariant gauge-independent solution to the problem that does not need the definition of a congruence of curves. Instead, starting from fundamental principles, in particular the parallel transport of 4-vectors, we derive the gravitational contribution to the frequency shift in a very natural way. It is in fact reasonable to expect that the gravitational effect should show up quite naturally and directly from the physical interpretation of the formulae. In doing so, we split the contribution of gravity from those effects that can be made zero under some reasonable assumptions or coordinate transformation --the fictitious forces.

This paper is structured as follows. We calculate the derivative of the frequency shift and make the parallel transport explicit in this formalism. We examine the terms arising from the differentiation in more detail, and we show that the gravitational effect comes from the derivative of the parallel transport. Finally, we discuss the limit for nearby geodesics, interesting for the application to GW detection.

\textit{The derivative of the frequency shift}-- Consider a thought experiment (see Fig.\;\ref{fig:diagram}) where a test mass, the \textit{emitter}, emits light beams toward another test mass, the \textit{receiver}. We shall consider only a single emitted beam and do a general relativistic calculation of the measured frequency shift.
The emission (reception) event is $\textbf{\e}\equiv x^\mu$ ($\textbf{\r}\equiv x^\mu$). The emitter (receiver) is moving with 4-velocity $u^\mu$ ($v^\mu$) under the action of an external non-gravitational 4-force per unit mass \footnote{Hereafter, 4-acceleration or (non-gravitational) 4-force are undistinguishable for the scopes of this work.} $f^\mu$ ($g^\mu$) and gravity. The proper time of the emitter (receiver) is $\tau_\ee$ ($\tau_\rr$). The light beam is characterised by the 4-momentum $k^\mu$ and the null-geodesic affine parameter $\lambda$.
We want to derive information about the underlying space-time curvature from the observed frequency shift.
It is worth noting that a phasemeter employed in a space-borne GW interferometer measures a phase difference between the incoming phase and the phase of a local laser beam \footnote{It is legitimate to assume that both the emitter and the receiver holds identical copies of the same laser oscillating at the reference frequency $\omega_\ee$.}, or the light may be split, directed toward different paths and then recombined as in a ground interferometer. In any case, the measured phase difference is related to the frequency difference via $\delta\omega=\d\delta\phi / \d t$. Therefore the natural observable for this calculation is the frequency difference between the received frequency $\omega_\rr$ and the emitted frequency $\omega_\ee$: $\delta\omega=\omega_\rr-\omega_\ee$. Now, the emitted (received) frequency is $\omega_\ee=k_\mu(\e)u^\mu(\e)$ [$\omega_\rr=k_\mu(\r)v^\mu(\r)$] \cite{misner,synge} and their difference can be written as follows
\begin{equation}\label{eq:freq_diff}
\delta\omega=k_\mu(\r)\left[v^\mu(\r)-u^\mu(\e)-\delta u^\mu\right],
\end{equation}
where $u^\mu$ has been subjected to a parallel transport from $\e$ to $\r$ along the light beam's null geodesic \cite{vitale2009}, i.e.\ $k^\alpha\nabla_\alpha u^\mu=0$. Here, compared to Ref.\ \cite{vitale2009}, the parallel transport has been made explicit such that $u^\mu(\r)=u^\mu(\e)+\delta u^\mu$,
where
\begin{equation}\label{eq:parallel_transport}
\delta u^\mu=\int_{\lambda_\ee}^{\lambda_\rr}\Gamma^\mu_{\alpha\beta}(\lambda)u^\alpha(\lambda) k^\beta(\lambda)\,\d\lambda
\end{equation}
can be found by integrating the parallel transport equation. Also, $\lambda_\rr=\lambda(\r)$ and $\lambda_\ee=\lambda(\e)$. Additionally, as $\r$ and $\e$ are causally connected by the beam's geodesic, it turns out that $\lambda$ and $\tau_\ee$ are in effect functions of $\tau_\rr$.
Differentiating Eq.\;\eqref{eq:freq_diff} with respect to $\tau_\rr$, the result is
\begin{widetext}
\begin{equation}
\label{eq:freq_diff_der}
\frac{\d\delta\omega}{\d\tau_\rr}=\underbrace{\frac{\text{D} k_\mu}{\d\tau_\rr} \left[v^\mu(\r)-u^\mu(\e)-\delta u^\mu\right]}_{\substack{\text{diff.\ velocity}\\ \text{+ par.\ transport}\\ \text{+ fictitious}\\ \textit{(a)}}}
+\underbrace{k_\mu(\r)\left[g^\mu(\r)-\tilde{f}^\mu(\e)\right]}_{\substack{\text{diff.\ force}\\ \text{+ fictitious}\\ \textit{(b)}}}
-\underbrace{k_\mu(\r)\frac{\text{D}\delta u^\mu}{\d\tau_\rr}}_{\substack{\text{par.\ transport deriv.}\\ \text{+ fictitious} \\ \textit{(c)}}}.
\end{equation}
\end{widetext}

\textit{Working out the differential-velocity term (a)}-- This term accounts for the difference of velocities, $v^\mu(\r)-u^\mu(\e)$, but also the parallel transport $\delta u^\mu$, both projected along $\text{D} k_\mu / \d\tau_\rr$. In the local Lorentz frame (LLF) at $\r$, $\Gamma^\alpha_{\mu\beta}=0$ so $\text{D} k_\mu / \d\tau_\rr \equiv \d k_\mu / \d\tau_\rr$ but, being an integral along a finite curve, $\delta u^\mu\neq0$. If the emitter and receiver are assumed to be infinitely close to each other such that the integral can be evaluated at a common reference frame coincident to the above-defined LLF at $\r$ (we shall later call it the LLF for \textit{nearby geodesics}) then additionally $\delta u^\mu=0$. Also, for every point along the null geodesic an LLF can also be defined such that the integrand in Eq.\;\eqref{eq:parallel_transport} will be zero locally. Therefore, while $\text{D} k_\mu / \d\tau_\rr$ includes local fictitious forces at $\r$, instead $\delta u^\mu$ reproduces the non-local effect coming from the integration of an infinitesimal fictitious force along the beam's null geodesic.

\textit{Working out the differential-force term (b)}-- Two contributions are included, the real force on the receiver, $g^\mu(\r)$, and an effective force on the emitter, $\tilde{f}^\mu(\e)$, which is given by
\begin{equation}
\tilde{f}^\mu=\frac{\d\tau_\ee}{\d\tau_\rr} f^\mu+\Gamma^\mu_{\alpha\beta}u^\alpha\left(v^\beta-\frac{\d\tau_\ee}{\d\tau_\rr} u^\beta\right),
\end{equation}
where $\d\tau_\ee/\d\tau_\rr=1+\delta\omega/\omega_\ee$. Because the non-gravitational forces are assumed small, it is safe to approximate $\d\tau_\ee/\d\tau_\rr\simeq1$ to first order. The effective force $\tilde{f}^\mu(\e)$ contains a $\Gamma^\mu_{\alpha\beta}$ term evaluated at $\e$ that, in general, is not zero when only the receiver's LLF is considered. However, this becomes zero in the nearby-geodesics LLF. Therefore this second term may be interpreted as a local fictitious force at the emitter.

\textit{Working out the parallel transport derivative term (c)}-- This is the gravitational contribution that the derivative of the frequency shift is sensitive to. Before going through the calculation, it is worth noting that $\Gamma^\mu_{\alpha\beta}$ may be still treated as a second rank tensor when $\alpha$ is fixed \cite{schutz}, i.e.\ $\Gamma^\mu_{\bar{\alpha}\beta}$ transforms as a tensor for fixed $\alpha\equiv\bar{\alpha}$, 
${\Gamma^\mu_{\bar{\alpha}\beta}}'=\Lambda^{\mu}_{~\zeta}\Lambda^{~\xi}_{\beta}\Gamma^\zeta_{\bar{\alpha}\xi}$, under a general transformation
 $\Lambda^{\mu}_{~\zeta}$ from the unprimed to the primed coordinate system. In differentiating the integral in $\delta u^\mu$ with respect to $\tau_\rr$, the extremes of integration depend on $\tau_\rr$, so the commutation between the derivative and the integral gives
\begin{equation}
\frac{\text{D}\delta u^\mu}{\d\tau_\rr}=\int_{\lambda_\ee}^{\lambda_\rr}v^\nu\nabla_\nu\left(\Gamma^\mu_{\alpha\beta}u^\alpha k^\beta\right)\d\lambda
+\left.\frac{\d\lambda}{\d\tau_\rr}\Gamma^\mu_{\alpha\beta} u^\alpha k^\beta\right|_\e^\r,
\end{equation}
where the integrand depends implicitly on $\lambda$. The last contribution becomes zero in the nearby-geodesics LLF, thus it acts as local inertial forces at $\e$ and $\r$. From now on, the focus is on rearranging the integral in a more familiar form. Applying the covariant derivative, the integrand may be recast as follows
\begin{equation}
\begin{split}
&v^\nu\nabla_\nu\left(\Gamma^\mu_{\alpha\beta}u^\alpha k^\beta\right) \\
&=\left(\partial_\nu\Gamma^\mu_{\alpha\beta}+\Gamma^\mu_{\xi\nu}\Gamma^\xi_{\alpha\beta}\right)u^\alpha v^\nu k^\beta 
+\Gamma^\mu_{\alpha\beta}\partial_\nu\left(u^\alpha  k^\beta\right)v^\nu.
\end{split}
\end{equation}
In effect, the quantity between parentheses reproduces the first two (positive) terms of the Riemann tensor. To make it more explicit, it is useful to consider a thought experiment where the role of the emitter and the receiver are reversed. The observed physical effect will be exactly the same. So, looking at Eq.\;\eqref{eq:freq_diff_der}, the role of the emitter and the receiver may be swapped, and in particular in the integral, $v^\nu\leftrightarrow u^\alpha$ and $\r\leftrightarrow \e$. The integral will formally be the same, but with reversed integration limits. Because of this symmetry, the original derivative of the parallel transport can be rewritten as
\begin{equation}
\begin{split}
\frac{\text{D}\delta u^\mu}{\d\tau_\rr}&=
\frac{1}{2}\int_{\lambda_\ee}^{\lambda_\rr}
R^\mu_{~\beta\nu\alpha}u^\alpha v^\nu k^\beta \\
&+\Gamma^\mu_{\alpha\beta}\nabla_\nu\left[k^\beta \left(u^\alpha v^\nu-u^\nu v^\alpha\right)
\right]\d\lambda.
\end{split}
\end{equation}
where the Riemann tensor arises from the antisymmetry between $\alpha$ and $\nu$.
The second term can be recast as $\Gamma^\mu_{\alpha\beta}\left[\nabla_\nu k^\beta \left(u^\alpha v^\nu-u^\nu v^\alpha\right)+k^\beta\mathcal{L}_\textbf{v}u^\alpha\right]$, where $\mathcal{L}_\textbf{v}u^\alpha=v^\nu\nabla_\nu u^\alpha-u^\nu\nabla_\nu v^\alpha=-\mathcal{L}_\textbf{u}v^\alpha$. It turns out that: (\textit{i}) it is zero in the LLF along the null geodesic, (\textit{ii}) it is zero when the two velocities are equal, and (\textit{iii}) it is zero in the nearby-geodesics LLF.
Integrated out along the beam's geodesic, it therefore gives a non-local fictitious force.
Instead, $R^\mu_{\;\nu\alpha\beta}\neq0$ can not be set to zero in a purposely chosen LLF: it is in fact
an intrinsic property of space-time that quantifies curvature. This curvature contribution is
\begin{equation}
\mathcal{R}^\mu = 
\frac{1}{2}\int_{\lambda_\ee}^{\lambda_\rr}
R^\mu_{~\nu\alpha\beta}u^\alpha v^\beta k^\nu\,\d\lambda.
\end{equation}
Under the general assumption that the non-gravitational forces are small, then the original formula becomes
\begin{widetext}
\begin{equation}
\label{eq:freq_diff_der_final}
\frac{\d\delta\omega}{\d\tau_\rr}=\frac{\text{D} k_\mu}{\d\tau_\rr} \left[v^\mu(\r)-u^\mu(\e)\right]
+k_\mu(\r)\left[g^\mu(\r)-f^\mu(\e)\right]
+k_\mu(\r)
\mathcal{R}^\mu
+\gamma_\text{fict}\left[\Gamma(\r),\Gamma(\e),\Gamma(\lambda)\right],
\end{equation}
\end{widetext}
where $\gamma_\text{fict}$
collects all additional terms that may depend on $\Gamma^\mu_{\alpha\beta}$ either locally (at event $\e$ or $\r$) or non-locally (via an integral along the beam's null geodesic). In any case, $\gamma_\text{fict}$ goes to zero in the nearby-geodesics LLF. The above result shows that not only the derivative of the frequency shift depends on the relative velocity and relative non-gravitational acceleration, but also contains a contribution from the integral of the Riemann tensor along the beam's geodesic, plus local and non-local fictitious forces.

\textit{The local reference frame for nearby geodesics}-- As an application, it is worth calculating the limit for nearby geodesics where a local \textit{approximately}-inertial reference frame common to both the receiver and the emitter can be set up.
In such conditions, $\d\tau_\rr\simeq\d\tau_\ee\simeq\d t$ so the effect of evaluating, for instance, a 4-vector at emission or reception is just a light time delay $\delta t=\delta x/c$, where $\delta x$ is the nominal distance between the emitter and the receiver. 
Additionally, $k^\mu=\omega_\ee/c\,n^\mu$, $\delta x^\mu=\delta x\,n^\mu$ and $n^\mu=(1,\hat{n})$. The Minkowski diagram of Fig.\;\ref{fig:diagram} shows this LLF and the geodesics of the emitter, receiver and light beam. In this reference frame the receiver appears still, and if we further assume that the emitter's apparent velocity is negligible to first order, then the derivative of the fractional frequency shift becomes
\begin{equation}
\begin{split}
\frac{\d}{\d t}\left(\frac{\delta\omega}{\omega_\ee}\right)&=\frac{1}{c}\frac{\d\hat{n}}{\d t}\cdot \left[\vec{v}(t)-\vec{u}(t-\delta t)\right] \\
&+\frac{1}{c}\hat{n}(t)\cdot[\vec{g}(t)-\vec{f}(t-\delta t)] \\
&+\frac{c}{2}\delta x
\left[R_{\mu\nu00}(t)-R_{\mu\nu00}(t-\delta t)\right] n^\mu\, n^\nu,
\end{split}
\end{equation}
where $\d\hat{n}/\d t=\vec{\Omega}\times\hat{n}$ and $\vec{\Omega}$ is the observed rotation of the line of sight \cite{congedo2013}. The additional fictitious forces contained in the more general result of Eq.\;\eqref{eq:freq_diff_der_final} are intentionally neglected here. As per the 4-velocity and 4-forces, the Riemann tensor must also be evaluated as a difference at the reception time and the delayed emission time. Moreover, all tensors are here evaluated in the defined LLF, and so is, of course, the Riemann tensor. So, if we want to relate this Riemann tensor to the one defined in the so-called transverse-traceless gauge \textit{wave frame} of an incoming GW, $R_{i0j0}^\text{\scalebox{0.8}{TT}}=1/2\,h_{ij,00}^\text{\scalebox{0.8}{TT}}$, a rotation $\Lambda^\mu_\nu$ will naturally show up transforming the wave frame into the LLF.

\textit{Conclusions}-- This paper has addressed the issue of intimately understanding how space-time curvature can affect the frequency shift of light beams exchanged between two test masses. The nature of the problem is general, but ultimately relevant to GW astronomy. Compared to previous work, the calculation is straightforward and does not require a congruence of curves, but is always matched to the ultimate physical interpretation.

Without assuming any coordinate system or fixing the gauge, thus retaining the full covariant nature of the formalism, we have found that the only gravitational contribution to the time derivative of the observed frequency shift comes just from the derivative of the parallel transport of the emitter's 4-velocity along the light beam's geodesic to the receiver. This is, in turn, given by an integral of the Riemann tensor along the null geodesic, although in a different form compared to previous work \cite{congedo2013,koop2014}. Additionally, fictitious forces show up locally at the receiver, at the emitter and non-locally integrated along the beam's geodesic. Those terms were not considered before, although we note here that they are all proportional to $\Gamma^\mu_{\alpha\beta}$, thus they are negligible under reasonable assumptions that are always met in practice.

As an application, we have calculated the derivative of the frequency shift for nearby geodesics in the reference frame of the receiver, showing that it is given in terms of differences between velocities, non-gravitational forces, and the Riemann tensor, all evaluated at time of reception and delayed time of emission. Additionally, the observed Riemann tensor can be directly related to the usual Riemann tensor defined in the wave frame where the gauge is transverse traceless.

It is worth commenting that all formulae containing the metric perturbation $h_{\mu\nu}$ to an underlying flat metric, as customary in GW astronomy (see e.g.\ Ref.\ \cite{vinet2013}), can be derived from the result contained in this paper by choosing a convenient coordinate system and by fixing the gauge. Nonetheless this does not alleviate, at least in principle, the problem of fictitious forces that may still be dependent on $h_{\mu\nu}$ itself. But, ultimately, the existence of a local LLF for both the receiver and the emitter \textit{justifies} the use of time-delayed differences in the calculation of the detector response to GWs. An example of which is found in Ref.\ \cite{congedo2015} where the detector response was given in terms of acceleration, a quantity that is very convenient for marginalising over detector systematics.

\begin{acknowledgments}
The author acknowledges support from the Beecroft Institute for Particle Astrophysics and Cosmology, and Oxford Martin School.
The author is very grateful to the following people for reading and commenting on this paper:
D.\ Alonso, T.\ Baker, P.\ G.\ Ferreira, H.\ A.\ Winther (Univ.\ of Oxford, UK); F.\ De Paolis (Univ.\ of Salento, Italy).
\end{acknowledgments}

\bibliography{references}

\end{document}